\documentclass[12pt]{article}
\usepackage{latexsym,amsfonts}
\makeatletter
\@addtoreset{equation}{subsection}
\makeatother

\begin{document}
\begin{center}
{\Large \bf General Relativity and Quantum Jumps:} \\[0.5cm]
{\large\bf The Existence of Nondiffeomorphic Solutions to the
Cauchy Problem in Nonempty Spacetime and Quantum Jumps as a
Provider of a Canonical Spacetime Structure}\\[1.5cm]
 {\bf Vladimir S.~MASHKEVICH}\footnote {E-mail:
  Vladimir\_Mashkevich@qc.edu}
\\[1.4cm] {\it Physics Department
 \\ Queens College\\ The City University of New York\\
 65-30 Kissena Boulevard\\ Flushing, New York
 11367-1519} \\[1.4cm] \vskip 1cm

{\large \bf Abstract}
\end{center}

It is shown that in spite of a generally accepted concept, there
exist nondiffeomorphic solutions to the Cauchy problem in nonempty
spacetime, which implies the necessity for canonical complementary
conditions. It is nonlocal quantum jumps that provide a canonical
global structure of spacetime manifold and, by the same token, the
canonical complementary conditions.

\newpage

\section*{Introduction}

In the literature, there exists no disagreement on the question of
the completeness of the Einstein equation. In all the books on
general relativity in which the completeness problem is considered
(see [1-13]), the latter is treated as resolved once and for all.
The reasoning is this. In the Cauchy problem, there are six
independent time evolution equations for six components $g_{ij}\;
(i,j=1,2,3)$ of metric, whereas the remaining four components
$g_{0\mu}\; (\mu =0,1,2,3)$ remain arbitrary. But diffeomorphic
metrics (ones connected by diffeomorphisms) are physically
equivalent, and a diffeomorphism involves exactly four degrees of
freedom. Hence it is concluded that all the solutions to the
Cauchy problem are diffeomorphic, i.e., equivalent. Therefore the
only task is to introduce four complementary conditions fixing a
solution. The complementary equations may be to a great extent
arbitrary, the only requirement is that thay be quasilinear. We
will call this resolution of the completeness problem a property
of diffeomorphic connectedness of the set of the solutions. For a
Ricci flat spacetime, diffeomorphic connectedness does take place
[14,15].

The Einstein equation is local, so that as long as it is complete,
general relativity is a local theory and does not determine any
global structure of spacetime manifold. This gives rise to the
problem of compatibility of general relativity with quantum
jumps, which are inherently nonlocal (see [16]).

In this paper, we argue that the coincidence of the number of
degrees of freedom of a diffeomorphism with the number of the
components $g_{0\mu}$ (or missing equations) does not imply
diffeomorphic connectedness for nonempty spacetime. Here is a
simple\, counterexample. Let $f$ be a function on a manifold $M^4$
and let $F:M^4\rightarrow M^4 $ be a diffeomorphism, so that
$(F^\ast f)(p)=f(Fp),\, p\in M^4 $. Let $\bar{f}$ be another
function. Is it true that there exists an $F$ such that
$\bar{f}=F^\ast f$\,? We have only one equation $\bar{f}(p)=f(Fp)$
and four degrees of freedom for $F$. But if the ranges of $\bar{f}$
and $f$ are different, ${\rm ran}\bar{f}\neq {\rm ran}f$, $F$ does
not exist. This counterexample may be straightforwardly extended
to the problem of diffeomorphic connectedness. Introduce
complementary conditions of the form $g_{0i}=0\;(i=1,2,3),\;R=f$
where $R$ is the scalar curvature and $f$ is a function. The
quasilinearity requirement is respected. If ${\rm ran}\bar{f}\neq
{\rm ran}f$, $\bar{g}$ and $g$ are not diffeomorphic.

The breakdown of the diffeomorphic connectedness implies the
necessity for canonical complementary conditions. It is quantum
jumps that provide the latter. Nonlocal quantum jumps click out a
universal cosmological time $t\in T$, so that spacetime manifold
has a canonical global structure: $M^4=T\times S$ where $S$ is a
cosmological space. The canonical complementary conditions are of
the form $dt=g(\partial/\partial t,\cdot)$, which corresponds to a
synchronous frame.

\section{The problem of the completeness of the Einstein equation
and diffeomorphic connectedness}

In this section, we summarize known facts and conventional
concepts on the problem of the completeness of the Einstein
equation and the Cauchy problem.

\subsection{The underdetermination of the Einstein equation}

The Einstein equations
\begin{equation}\label{1.1.1}
G^\nu_\mu-\Lambda g^\nu_\mu=T^\nu_\mu\,,\qquad \mu,\nu=0,1,2,3
\end{equation}
form a system of ten equations in the ten metric components
$g_{\mu\nu}$ and their first and second derivatives. However the
covariant divergence of each side vanishes identically, i.e.,
\begin{equation}\label{1.1.2}
\left(G^\nu_\mu-\Lambda g^\nu_\mu\right)_{;\nu}=0
\end{equation}
and
\begin{equation}\label{1.1.3}
T^\nu_{\mu;\nu}=0
\end{equation}
hold independently of the field (metric and matter) equations.
Thus (1.1.1) provides only six independent equations for the ten
components $g_{\mu\nu}$.

\subsection{The underdetermination of the Cauchy problem}

In the Cauchy problem, the equations of time evolution are
\begin{equation}\label{1.2.1}
G^i_j-\Lambda g^i_j=T^i_j\,,\qquad i,j=1,2,3
\end{equation}
The equations
\begin{equation}\label{1.2.2}
G^0_\mu-\Lambda g^0_\mu=T^0_\mu\,,\qquad \mu=0,1,2,3
\end{equation}
are constraints on initial conditions for $g_{\mu\nu}$ and
\begin{equation}\label{1.2.3}
\dot{g}_{ij}=g_{ij,0}=\partial g_{ij}/\partial t\,,\quad t=x^0
\end{equation}
Thus there are only six dynamical equations for the ten components
$g_{\mu\nu}$, so that a solution to the Cauchy problem is
determined up to four degrees of freedom, i.e., four arbitrary
functions.

\subsection{Diffeomorphic connectedness}

It is the commonly accepted concept that the set of all solutions
to the Cauchy problem possesses the property of diffeomorphic
connectedness: for any two solutions, $g$ and $\bar{g}$, there
exists a diffeomorphism $F:M^4\rightarrow M^4$ such that
$\bar{g}=F^\ast g$. Since diffeomorphic metrics describe the same
physical situation, the latter is uniquely determined in the
Cauchy problem. The ground for the diffeomorphic connectedness is
this: There are four degrees of freedom in the solution, and a
diffeomorphism involves exactly four degrees of freedom; thus the
solution is determined up to a diffeomorphism.

For a Ricci flat spacetime, diffeomorphic connectedness  does take
place [14,15].

\subsection{Complementary conditions}

As long as diffeomorphic connectedness takes place, i.e., all the
solutions to the Cauchy problem are physically equivalent, the
only task is to introduce four complementary conditions fixing a
solution.

In the literature, a variety of the complementary conditions is
presented: harmonic coordinates [1,5]; lapse and shift functions
(ADM formalism) [2]; synchronous coordinates [9,6]; equations for
$\ddot{g}_{0\mu}$ [8]. All those are noncovariant and exploit a
coordinate system.

\section{The invalidity of diffeomorphic connectedness\\ in nonempty spacetime}

In this section, we introduce a counterexample to show that in
nonempty spacetime diffeomorphic connectedness is not valid.

\subsection{A time-space hypoframe and frame}

As long as it is possible, it is reasonable to use a
coordinate-free approach. For this purpose, we introduce the
following objects:

a timelike vector field $b_0$ and a 1-form $\beta$ such that
$\beta (b_0)>0$, or $\beta=f_0 \beta ^0,\;\beta ^0(b_0)=1$ and a
function $f_0>0$;

a frame $(b_a:a=0,1,2,3)$ such that $\beta ^0(b_a)=\delta ^0_a$.

We will call $(b_0,\beta)$ a time-space hypoframe and $(b_a)$ a
time-space frame.

\subsection{A synchronous hypoframe and four noncovariant
conditions}

A time-space hypoframe $(b_0,\beta^0)$ will be called a
synchronous hypoframe. It provides a means of introducing four
noncovariant conditions
\begin{equation}\label{2.2.1}
g(b_0,\cdot)=\beta^0
\end{equation}
or
\begin{equation}\label{2.2.2}
g(b_0,b_a)=\delta_{0a}\,,\quad a=0,1,2,3
\end{equation}

\subsection{Scalar curvature invariant condition}

Consider a nonempty, i.e., not a Ricci flat spacetime. Introduce a
complementary condition of the form
\begin{equation}\label{2.3.1}
R=f_R
\end{equation}
where $f_R$ is a function on $M^4$. This condition is not only
coordinate-free but invariant as well. It plays a crucial role in
the argumentation that follows.

\subsection{(3+1) conditions and a holonomic frame}

In addition to (2.3.1), introduce three noncovariant conditions of
the form
\begin{equation}\label{2.4.1}
g(b_0,b_j)=0\,,\quad j=1,2,3
\end{equation}
where $(b_a)$ is a time-space frame. We will call conditions
(2.4.1) and (2.3.1) (3+1) conditions.

Now we introduce a holonomic frame:
\begin{equation}\label{2.4.2}
(b_a)\rightarrow (b_\mu)\,,\quad
b_\mu=\partial_\mu=\partial/\partial x^\mu
\end{equation}
where $x=(x^\mu)$ are (local) coordinates. Now the (3+1) conditions
take the form
\begin{equation}\label{2.4.3}
g_{0j}(x)=0\,,\quad j=1,2,3
\end{equation}
\begin{equation}\label{2.4.4}
R(x)=f_R(x)
\end{equation}

In the case that all the four conditions are noncovariant
(equations (2.2.2) are an important example), all the four degrees
of freedom in metric may be described by coordinate functions
$x^\mu=x^\mu(p),\; p\in M^4$. If one of the conditions is
invariant, those functions describe only three degrees of freedom;
one degree of freedom is described by a function involved in the
invariant condition. The only invariant condition is (2.3.1) or
(2.4.4) since $R$ is the only first order invariant. Thus it is
the function $f_R$ that describes one degree of freedom.

\subsection{A system of quasilinear differential equations\\
for metric components}

The complementary equations may be to a great extent arbitrary.
The only requirement is that these equations, like the Einstein
equations (1.2.1), be quasilinear. Then there will be a complete
system of quasilinear differential equations for metric
components. Let us verify that the (3+1) conditions do respect that
requirement.

The Ricci tensor
\begin{equation}\label{2.5.1}
R^\sigma_\mu=g^{\sigma\kappa}R_{\kappa\mu}\;,\quad
R_{\kappa\mu}=g^{\eta\lambda}R_{\eta\kappa\lambda\mu}
\end{equation}
the Riemann tensor
\begin{equation}\label{2.5.2}
R_{\eta\kappa\lambda\mu}=\frac{1}{2}
\left(g_{\eta\mu,\kappa\lambda}+g_{\kappa\lambda,\eta\mu}
-g_{\eta\lambda,\kappa\mu}-g_{\kappa\mu,\eta\lambda}\right)
+\left(g^{\nu\rho}\Gamma_{\nu\kappa\lambda}\Gamma_{\rho\eta\mu}-
g^{\nu\rho}\Gamma_{\nu\kappa\mu}\Gamma_{\rho\eta\lambda}\right)
\end{equation}
the Christoffel symbol
\begin{equation}\label{2.5.3}
\Gamma_{\eta\kappa\lambda}=\frac{1}{2}\left(g_{\eta\kappa,\lambda}+
g_{\eta\lambda,\kappa}-g_{\kappa\lambda,\eta}\right)
\end{equation}
From noncovariant conditions (2.4.3)  it follows that
\begin{equation}\label{2.5.4}
g_{0j}=0,\quad g^{0j}=0,\quad g^{00}=1/g_{00}
\end{equation}
We find
\begin{equation}\label{2.5.5}
R^p_m=-\frac{1}{2}g^{00}g^{pk}g_{km,00}+
\frac{1}{4}\left(g^{00}\right)^2g^{pk}g_{km,0}g_{00,0}+\tilde{R}^p_m
\end{equation}
\begin{equation}\label{2.5.6}
R^0_0=-\frac{1}{2}g^{00}g^{km}g_{km,00}+
\frac{1}{4}\left(g^{00}\right)^2g^{km}g_{km,0}g_{00,0}+\tilde{R}^0_0
\end{equation}
where
\begin{equation}\label{2.5.7}
\tilde{R}^0_0=g^{00}g^{km}\tilde{R}_{0k0m}\; ,\quad
\tilde{R}^p_m=g^{00}g^{pk}\tilde{R}_{0k0m}+g^{pk}g^{il}\tilde{R}_{iklm}
\end{equation}
\begin{equation}\label{2.5.8}
\tilde{R}_{0k0m}=-\frac{1}{2}g_{00,km}+\frac{1}{4}g^{00}g_{00,k}g_{00,m}+
\frac{1}{4}g^{nr}g_{nk,0}g_{rm,0}+\frac{1}{2}g^{nr}\Gamma_{nkm}g_{00,r}
\end{equation}
\begin{equation}\label{2.5.9}
\begin{array}{l}
\tilde{R}_{iklm}=R_{iklm}\\{\displaystyle=\frac{1}{2}\left(g_{im,kl}+g_{kl,im}-
g_{il.km}-g_{km,il}\right)+
\left(g^{nr}\Gamma_{nkl}\Gamma_{rim}-g^{nr}\Gamma_{nkm}\Gamma_{ril}\right)}\\
\quad {}
{\displaystyle+\frac{1}{4}g^{00}\left(g_{kl,0}g_{im,0}-g_{km,0}g_{il,0}\right)}
\end{array}
\end{equation}
Finally,
\begin{equation}\label{2.5.10}
R=-g^{00}g^{mk}g_{km,00}+\frac{1}{2}(g^{00})^2
g^{mk}g_{km,0}g_{00,0}+\tilde{R}
\end{equation}
where
\begin{equation}\label{2.5.11}
\tilde{R}=\tilde{R}^\mu_\mu=\tilde{R}^0_0+\tilde{R}^m_m
\end{equation}

The quantities $\tilde{R}^p_m$ and $\tilde{R}$ involve
$g_{ij}\,,\;g_{ij,k}\,,\;g_{ij,kl}\,,\;g_{ij,0}\,;
\;g_{00}\,,\;g_{00,i}\,,\;g_{00,ij}$\,, i.e., in fact,
$g_{ij}\,,\;g_{ij,0}\,;\;g_{00}$\,. Thus, in view of (2.5.5) and
(2.5.10), the seven equations
\begin{equation}\label{2.5.12}
R^i_j-\frac{1}{2}\delta^i_jR-\delta^i_j\Lambda=T^i_j\,,\quad R=f_R
\end{equation}
form a system of quasilinear equations for the seven metric
components $g_{ij}\,,\;g_{00}$\,, the higher time derivatives
being $\ddot{g}_{ij}$ and $\dot{g}_{00}$\,, respectively. Initial
conditions are those for $g_{ij}\,,\;\dot{g}_{ij}$ and $g_{00}$
obeying constraints (1.2.2).

\subsection{The existence of nondiffeomorphic solutions\\
to the Cauchy problem in nonempty spacetime}

Let $g$ and $\bar{g}$ be solutions to the system of equations
(2.5.12) with functions $f_R$ and $\bar{f}_R$, respectively. If
the ranges of the functions are different,
\begin{equation}\label{2.6.1}
{\rm ran}\bar{f}_R\neq {\rm ran}f_R
\end{equation}
$g$ and $\bar{g}$ are nondiffeomorphic. Indeed, $\bar{g}=F^\ast g$
implies $\bar{R}=F^\ast R$ so that
\begin{equation}\label{2.6.2}
{\rm ran}\bar{R}={\rm ran}R
\end{equation}
which does not hold. Thus the set of all the solutions to the
Cauchy problem in nonempty spacetime is not diffeomorphically
connected. It is the degree of freedom described by the function
$f_R$ that brings about breaking diffeomorphic connectedness.

It is necessary to point out the following. From (1.1.1) follows
\begin{equation}\label{2.6.3}
R=-T+4\Lambda
\end{equation}
where $T=T^\mu_\mu$. Therefore if matter dynamics had been
independent of metric, it would have been impossible to introduce
an arbitrary function $f_R$ since the scalar curvature would have
been prescribed. But that independence holds only in an empty
spacetime.

In a nonempty spacetime, (2.6.3) and (2.3.1) are two independent
equations for metric and matter.

For example, in the case of one scalar matter field $\varphi$, we
have eight functions: $(g_{ij}),\,g_{00}$ and $\varphi$, and eight
equations: (1.2.1), (2.3.1) and the wave equation
\begin{equation}\label{2.6.4}
\Box \varphi+m^2\varphi+\xi R\varphi=0
\end{equation}

\subsection{The misleadingness of infinitesimal diffeomorphisms}

One must not be misled by infinitesimal diffeomorphisms.

Let $g$ and $\bar{g}=g+h$ with $h$ infinitesimal be two solutions
to the Cauchy problem. We will show that there exists an
infinitesimal local diffeomorphism $F$ such that
\begin{equation}\label{2.7.1}
\bar{g}=F^\ast g
\end{equation}

In (local) coordinates $(x^\mu)$ we have
\begin{equation}\label{2.7.2}
g_{\mu\nu}+h_{\mu\nu}=(F^\ast g)_{\mu\nu}
\end{equation}
or
\begin{equation}\label{2.7.3}
g_{\mu\nu}+h_{\mu\nu}=g(F_\ast\partial_\mu\,,F_\ast\partial_\nu)
\end{equation}
For an infinitesimal $F$, we have
\begin{equation}\label{2.7.4}
F_\ast\partial_\mu=\partial_\mu+L_{v^\lambda\partial_\lambda}
\partial_\mu=\partial_\mu+[v^\lambda\partial_\lambda,\partial_\mu]=
\partial_\mu-v^\lambda{}_{,\mu}\partial_\lambda
\end{equation}
where $v^\lambda\partial_\lambda$ is an infinitesimal vector
field. Thus we obtain
\begin{equation}\label{2.7.5}
g_{\mu\lambda}v^\lambda{}_{,\nu}+g_{\nu\lambda}v^\lambda{}_{,\mu}=
-h_{\mu\nu}
\end{equation}
or
\begin{equation}\label{2.7.6}
{\rm for}\quad\mu\nu=00\qquad 2g_{0\lambda}v^\lambda{}_{,0}=-h_{00}
\end{equation}
\begin{equation}\label{2.7.7}
{\rm for}\quad\mu\nu=0j\qquad g_{0\lambda}v^\lambda{}_{,j}+
g_{j\lambda}v^\lambda{}_{,0}=-h_{0j}
\end{equation}
\begin{equation}\label{2.7.8}
{\rm for}\quad\mu\nu=ij\qquad g_{i\lambda}v^\lambda{}_{,j}+
g_{j\lambda}v^\lambda{}_{,i}=-h_{ij}
\end{equation}
Let $(x^\mu)$ be synchronous coordinates for $g$,
\begin{equation}\label{2.7.9}
g_{0\mu}=\delta_{0\mu}
\end{equation}
then
\begin{equation}\label{2.7.10}
v^0{}_{,0}=-\frac{1}{2}h_{00}
\end{equation}
\begin{equation}\label{2.7.11}
g_{ji}v^i{}_{,0}=-h_{0j}-v^0{}_{,j}\,,\qquad
v^i{}_{,0}=-g^{ij}\left(h_{0j}+v^0{}_{,j}\right)
\end{equation}
\begin{equation}\label{2.7.12}
h_{ij}=-\left(g_{il}v^l{}_{,j}+g_{jl}v^l{}_{,i}\right)
\end{equation}
whence
\begin{equation}\label{2.7.13}
v^0(t,\vec{x})=v^0(0,\vec{x})-\frac{1}{2}\int\limits_0^t
h_{00}(t',\vec{x})dt'
\end{equation}
\begin{equation}\label{2.7.14}
v^i(t,\vec{x})=v^i(0,\vec{x})-
\int\limits_0^t[g^{ij}(h_{0j}+v^0{}_{,j})](t',\vec{x})dt'
\end{equation}
Initial conditions are
\begin{equation}\label{2.7.15}
h(0,\vec{x})=0\,,\; {\rm i.e.}, \quad h_{0\mu}(0,\vec{x})=0,\quad
h_{ij}(0,\vec{x})=0
\end{equation}
\begin{equation}\label{2.7.16}
h_{ij,0}(0,\vec{x})=0
\end{equation}
The components $h_{0\mu}(t,\vec{x})$ are given with
$h_{0\mu}(0,\vec{x})=0$.

Put
\begin{equation}\label{2.7.17}
v^\mu(0,\vec{x})=0
\end{equation}
then
\begin{equation}\label{2.7.18}
v^\mu{}_{,l}(0,\vec{x})=0
\end{equation}
and
\begin{equation}\label{2.7.19}
v^l{}_{,j0}=v^l{}_{,0j}=-[g^{lk}(h_{0k}+v^0{}_{,k})]_{,j}=0 \quad{\rm
for}\quad x=(0,\vec{x})
\end{equation}
From (2.7.12), (2.7.18), (2.7.19) follows
\begin{equation}\label{2.1.20}
h_{ij}(0,\vec{x})=0,\quad h_{ij,0}(0,\vec{x})=0
\end{equation}
Thus the formulas
\begin{equation}\label{2.7.21}
v^0(t,\vec{x})=-\frac{1}{2}\int\limits^t_0 h_{00}(t',\vec{x})dt'
\end{equation}
\begin{equation}\label{2.7.22}
v^i(t,\vec{x})=-\int\limits_0^t
[g^{ij}(h_{0j}+v^0{}_{,j})](t',\vec{x})dt'
\end{equation}
give $F_\ast\partial_\mu$ (2.7.4).

Now put in the (3+1) conditions
\begin{equation}\label{2.7.23}
\bar{f}_R=f_R+w
\end{equation}
with $w$ infinitesimal such that (2.6.1) is fulfilled; e.g.,
\begin{equation}\label{2.7.24}
{\rm ran}f_R=[a,\infty),\quad {\rm ran}\bar{f}_R=[\bar{a},\infty),
\quad \bar{a}=a+c,\quad c\neq0,\quad c\;\;{\rm infinitesimal}
\end{equation}
Then $\bar{g}$ and $g$ are not diffeomorphic. On the other hand,
$\bar{g}=g+h$ with $h$ infinitesimal so that infinitesimally
$\bar{g}=F^\ast g$ whence $\bar{R}=F^\ast R$, i.e.,
$\bar{f}_R=F^\ast f_R$ in spite of (2.6.1).

Again, $f$ and $\bar{f}=f+w$ with $w$ infinitesimal are, in
general, infinitesimally diffeomorphic. The equality
\begin{equation}\label{2.7.25}
\bar{f}=F^\ast f
\end{equation}
means
\begin{equation}\label{2.7.26}
f(x)+w(x)=f(Fx)
\end{equation}
Put
\begin{equation}\label{2.7.27}
Fx=x+\xi(x)
\end{equation}
so that
\begin{equation}\label{2.7.28}
f(x)+w(x)=f(x+\xi(x))
\end{equation}
Infinitesimally
\begin{equation}\label{2.7.29}
f(x+\xi(x))=f(x)+[(\partial_\mu f)\xi^\mu](x)
\end{equation}
whence
\begin{equation}\label{2.7.30}
(\partial_\mu f)\xi^\mu=w
\end{equation}
The solvability criterion is $\partial_\mu f\neq0$.

The consideration carried out demonstrates that an infinitesimal
(local) diffeomorphism may be misleading.

\section{Quantum jumps and a canonical spacetime manifold
structure}

In this section, we show that quantum jumps provide a canonical
structure of spacetime manifold and, by the same token, canonical
complementary conditions.

\subsection{The necessity for canonical complementary conditions}

Had diffeomorphic connectedness taken place, all complementary
conditions would have been equivalent, and choosing some of them
would have been a matter of convenience. The breakdown of the
connectedness necessitates the introduction of specific, canonical
conditions. When introducing the latter, we should be guided by both
mathematical and physical reasons.

\subsection{Synchronous conditions}

We restrict our choice to global coordinate-free complementary
conditions involving no arbitrary functions. Such conditions are
provided by a synchronous hypoframe
\begin{equation}\label{3.2.1}
(b_0,\beta^0),\quad \beta^0(b_0)=1
\end{equation}
and are of the form of (2.2.1):
\begin{equation}\label{3.2.2}
g(b_0\,,\cdot)=\beta^0
\end{equation}
We call this conditions synchronous since in a holonomic frame
\begin{equation}\label{3.2.3}
b_0\rightarrow\partial/\partial t,\quad
b_j\rightarrow\partial/\partial x^j
\end{equation}
they take the form of (2.7.9), which corresponds to synchronous
coordinates.

There is a family of synchronous hypoframes and, by the same
token, a family of synchronous conditions. A synchronous hypoframe
(3.2.1) has seven degrees of freedom, in view of which we will not
raise the question on diffeomorphic connectedness. Our goal is to
introduce a specific, canonical hypoframe.

\subsection{Semiclassical gravity and quantum jumps}

It is quantum jumps that, due to their nonlocality, provide a
canonical global structure of spacetime manifold and, by this
structure, a canonical hypoframe.

In semiclassical gravity, the energy-momentum tensor
\begin{equation}\label{3.3.1}
T=\left(\Psi,\hat{T}\Psi\right)
\end{equation}
where $\hat{T}$ is the energy-momentum tensor operator and $\Psi$
is a state vector. A quantum jump is that of the state vector
(from here on see [17]):
\begin{equation}\label{3.3.2}
\Psi_{\rm befote\; jump}=:\Psi^<\rightarrow \Psi^>:=\Psi_{\rm
after\; jump}
\end{equation}
A jump of $\Psi$ results in that of $T$:
\begin{equation}\label{3.3.3}
\Delta
T=\left(\Psi^>,\hat{T}\Psi^>\right)-\left(\Psi^<,\hat{T}\Psi^<\right)
\end{equation}
under the assumption that $\hat{T}$ is continuous. Discontinuity
of $T$ causes a violation of the Einstein equation (1.1.1). The
components $G^i_j$ involve the second time derivatives
$\ddot{g}_{ij}$, which makes it possible to retain the six
equations (1.2.1). Jumps of $T_j^i$ will result in those of
$\ddot{g}_{ij}$.

\subsection{A canonical spacetime manifold structure}

A quantum jump of the state vector gives rise to a set of
events---jumps of $\ddot{g}_{ij}$. Those events are, by
definition, simultaneous, which allows for synchronizing clocks
and thereby furnishing a universal time. The latter, in its turn,
implies the product spacetime manifold:
\begin{equation}\label{3.4.1}
M=M^4=T\times S,\quad M\ni p=(t,s),\quad t\in T,\quad-\infty\leq
t_{\rm min}\leq t\leq t_{\rm max}\leq\infty,\quad s\in S
\end{equation}
The one-dimensional manifold $T$ is the universal cosmological time,
the three-dimensional manifold $S$ is a cosmological space. By
(3.4.1), the tangent space $M_p$ at a point $p\in M$ is
\begin{equation}\label{3.4.2}
M_p=T_t\oplus S_s\,,\quad p=(t,s)
\end{equation}

In special relativity, the concept of simultaneity relating to
quantum jumps makes no operationalistic sense. Taking gravity into
account endows the concept with an operationalistic content---the
simultaneity of the jumps of $\ddot{g}_{ij}$. On the other hand,
it is the simultaneity related to quantum jumps that provides the
global structure (3.4.1) for spacetime manifold. General
relativity per se is a local theory: ``Indeed general relativity
does not prescribe the topology of the world \ldots'' (Weyl [18]).
Thus general relativity and quantum jumps complement each other.

\subsection{A canonical synchronous hypoframe and canonical
complementary conditions}

We introduce a projection function on $M^4$:
\begin{equation}\label{3.5.1}
t:M^4\rightarrow T,\quad p=(s,t)\mapsto t(p)=t
\end{equation}
Now we put
\begin{equation}\label{3.5.2}
\beta^0=dt,\quad b_0=\partial/\partial t
\end{equation}
($t$ is a coordinate function). For any frame $(b_j)$ on $S$, we
have
\begin{equation}\label{3.5.3}
\beta^0(b_a)=\delta_a^0
\end{equation}
The hypoframe
\begin{equation}\label{3.5.4}
(b_0,\beta^0)=(\partial/\partial t,dt)
\end{equation}
is the canonical one.

The canonical complementary conditions take the form
\begin{equation}\label{3.5.5}
dt=g(\partial/\partial t,\cdot)
\end{equation}
In view of the global character of time $t$, these conditions are
global and, in fact, coordinate-free. We may rewrite (3.5.5) in
the explicitly coordinate-free form:
\begin{equation}\label{3.5.6}
\beta^0=g(b_0,\cdot)
\end{equation}
In connection with this, we quote Weyl [18]: ``The introduction of
numbers as coordinates \ldots is an act of violence whose only
practical vindication is the special calculatory manageability of
the ordinary number continuum with its four basic operations.''

By (3.5.5), metric is of the form
\begin{equation}\label{3.5.7}
g=dt\otimes dt-h_t
\end{equation}
where $h_t$ is a Riemannian metric on $S$ depending on $t$. Thus
we have
\begin{equation}\label{3.5.8}
T_t\bot S_s\,,\quad (t,s)\in T\times S
\end{equation}

The problem of the underdetermination of the Einstein equation is
resolved by the canonical complementary conditions, which are
provided by nonlocal quantum jumps. Thus, quantum jump nonlocality
not only does not contradict relativity, but it is essential for
general relativity to be a complete theory. Quantum jumps occur in
nonempty spacetime---just where the underdetermination problem
arises.

\subsection{A canonical decomposition of the energy-momentum
tensor}

The canonical structure of spacetime manifold implies a canonical
decomposition of the energy-momentum tensor:
\begin{equation}\label{3.6.1}
T=T_E+T_P+T_S
\end{equation}
\begin{equation}\label{3.6.2}
T_{E\mu\nu}=\delta_{\mu 0}\delta_{\nu 0}T_{00},\quad
T_{P\mu\nu}=\delta_{\mu 0}\delta_\nu^j T_{0j}+\delta_{\nu
0}\delta_\mu^j T_{0j},\quad T_{S\mu\nu}=\delta_\mu^i \delta_\nu^j
T_{ij}
\end{equation}
$T_E$, $T_P$, and $T_S$ are the energy, momentum, and stress tensors,
respectively. We have correspondences:
\begin{equation}\label{3.6.3}
T_E\leftrightarrow T_{00},\quad T_P\leftrightarrow
\vec{K},\quad K_j=T_{0j}
\end{equation}
$T_{00}$ and $\vec{K}$ are energy and momentum
densities, respectively. Energy
\begin{equation}\label{3.6.4}
E=\int\limits_S d\vec{x}\sqrt{|h|}\,T_{00}
\end{equation}
momentum
\begin{equation}\label{3.6.5}
\vec{P}=\int\limits_S d\vec{x}\sqrt{|h|}\,
\vec{K}
\end{equation}
where $|h|=\rm{det}(h_{ij})$.

\section*{Acknowledgments}

I would like to thank Alex A. Lisyansky for support and Stefan V.
Mashkevich for helpful discussions.

\end{document}